\documentclass[tinyml,screen]{acmart}

\usepackage{url}            % simple URL typesetting
\usepackage{booktabs}       % professional-quality tables
\usepackage{multirow}
\usepackage{nicefrac}       % compact symbols for 1/2, etc.
\usepackage{microtype}      % microtypography
\usepackage{amsmath,amsfonts} %amssymb
\usepackage{pifont}         % for tick and cross symbols
\usepackage{gensymb}        % for degree symbol
\usepackage{graphicx}
\usepackage{textcomp}
\usepackage{xcolor}
\usepackage{tabularx}
\usepackage{subcaption}
\usepackage{enumitem}
\usepackage{siunitx}
\usepackage{stfloats}
\usepackage[ruled,vlined]{algorithm2e}
\usepackage{color, colortbl}

\AtBeginDocument{%
  \providecommand\BibTeX{{%
    \normalfont B\kern-0.5em{\scshape i\kern-0.25em b}\kern-0.8em\TeX}}}

\setcopyright{rightsretained}
\copyrightyear{2022}
\acmYear{2022}

\begin{document}

% \title{Runtime Allocation of Harvested Energy using Reinforcement Learning for Energy-Neutral IoT Devices}
\title[tinyMAN: Energy Management Framework using Reinforcement Learning]{tinyMAN: Lightweight Energy Manager using Reinforcement Learning for Energy Harvesting Wearable IoT Devices}

\author{Toygun Basaklar}
\email{basaklar@wisc.edu}
\affiliation{%
  \institution{University of Wisconsin-Madison}
  \city{Madison}
  \state{Wisconsin}
  \country{USA}
  \postcode{53706}
}

\author{Yigit Tuncel}
\email{tuncel@wisc.edu}
\affiliation{%
  \institution{University of Wisconsin-Madison}
  \city{Madison}
  \state{Wisconsin}
  \country{USA}
  \postcode{53706}
}

\author{Umit Y. Ogras}
\email{uogras@wisc.edu}
\affiliation{%
  \institution{University of Wisconsin-Madison}
  \city{Madison}
  \state{Wisconsin}
  \country{USA}
  \postcode{53706}
}

% \renewcommand{\shortauthors}{Basaklar, et al.}

\begin{abstract}

%\thesis{Increasing health awareness and advancements in Internet of Things (IoT) devices have resulted in a significant growth in wearable edge devices and applications for remote health monitoring.}
% Increase in health awareness and advancements in Internet of Things (IoT) devices have resulted in a significant growth in wearable edge devices and applications for remote health monitoring. 
% For example, assistive devices for Parkinson's Disease patients need to provide precisely timed audio cues to cope with gait disturbances~\cite{deb2021trends,ginis2018cueing}.
% These devices must operate at a tight energy and computational budget ($\sim \hspace{-1mm}\mu$W) due to limited battery capacity~\cite{bhat2020self,de2017feasibility}. 
Advances in low-power electronics and machine learning techniques lead to many novel wearable IoT devices.
These devices have limited battery capacity and computational power.
Thus, energy harvesting from ambient sources is a promising solution to power these low-energy wearable devices.
They need to manage the harvested energy optimally to achieve energy-neutral operation, which eliminates recharging requirements.
Optimal energy management is a challenging task due to the dynamic nature of the harvested energy and the battery energy constraints of the target device.
To address this challenge, we present a reinforcement learning based energy management framework, tinyMAN, for resource-constrained wearable IoT devices.
The framework maximizes the utilization of the target device under dynamic energy harvesting patterns and battery constraints.
Moreover, tinyMAN does not rely on forecasts of the harvested energy which makes it a prediction-free approach.
We deployed tinyMAN on a wearable device prototype using TensorFlow Lite for Micro thanks to its small memory footprint of less than 100 KB.
Our evaluations show that tinyMAN achieves less than 2.36 ms and 27.75 $\upmu$J while maintaining up to 45\% higher utility compared to prior approaches. 
% Measurements on this device shows that \todo{tinyMAN} has \todo{$Y\times$} smaller energy overhead than iterative approaches. 

% \todo{$Y\times$} smaller energy overhead than the iterative approach. 
% \todo{XX power consumption and XX inference time.}

% EH alone is not sufficient to achieve energy-neutral operation.
% Energy neutrality can enable self-sustainability and automated battery charging.
% Achieving energy neutrality is a challenging task due to the dynamic nature of the harvested energy.

% It takes current and initial battery levels, and the harvested energy at previous time interval as inputs and outputs an energy allocation value.
% \todo{tinyMAN} is trained using in-house developed environment for RL which makes use of the solar and motion EH modalities and \textit{American Time Use Survey} data from 4772 different users. 
% Our framework achieves up to \todo{xx\%} higher utility compared to the state-of-the-art solutions while staying within \todo{xx\%} of the iterative optimal solution.
% In addition, \todo{tinyMAN} is deployed on a wearable device prototype using TensorFlow Lite for Microcontrollers and achieves \todo{$Y\times$} smaller energy overhead than the iterative approach. 
% \todo{XX power consumption and XX inference time.} 

\end{abstract}

 %, model compression,learning at the edge}

\maketitle

\section{Introduction}
\label{sec:introduction}

The emergence of small form-factor and low-cost wearable Internet of Things (IoT) devices lead to many novel edge-computing use cases~\cite{capra2019edge,wang2020convergence,bianchi2019iot}.
%The emergence of the Internet of Things (IoT) devices~\cite{capra2019edge,wang2020convergence} and the reduction of their size and cost~\cite{bianchi2019iot} lead to many novel wearable IoT devices.
% Advances in low-power electronics and machine learning techniques lead to many novel Internet of Things (IoT) devices that can be deployed at the edge. 
% Wearable IoT devices fuse data from multiple sensors, such as inertial measurement units (IMU) and biopotential amplifiers, to achieve accurate real-time tracking.
These include promising applications at the edge ranging from remote health monitoring to smart livestock monitoring systems~\cite{hiremath2014wearable,yamin2021online,tuncel2021much, lau2019survey}.
The devices that run these applications must operate with a tight energy budget ($\sim \hspace{-1mm}\upmu$W) and computational power due to limited battery capacity and small form-factor to be practical~\cite{de2017feasibility, basaklar2021hypervector}.%bhat2020self
% Edge devices typically have limited battery size due to small form-factor, especially in wearable devices. 
The small battery capacity limits the battery lifetime and requires frequent recharging, deteriorating the user experience.
To mitigate this effect, energy harvesting (EH) from ambient sources, such as light, motion, electromagnetic waves, and body heat, has emerged as a promising solution to power these devices~\cite{kansal2007power,tuncel2020towards}.%tuncel2020towards.

Energy-neutral operation (ENO) is achieved if the total energy consumed over a given period equals the energy harvested in the same period.
EH solutions should achieve ENO to ensure that the device maintains a certain battery level by continuously recharging the battery.
%itself battery charging becomes automated.
However, relying only on EH is not sufficient to achieve energy neutrality due to the uncertainties of ambient sources.
The application performance and utilization of the device can tank in low EH conditions~\cite{fraternali2020ember}.
%as they stop allocating energy . 
Energy management algorithms need to use the available energy judiciously to maximize the application performance while minimizing manual recharge interventions to tackle this challenge~\cite{tuncel2021eco}. 
% EH prediction mechanism is needed such that the algorithm spares available energy for minimum level of operation which is also a challenging task depending on the dynamic resource demand of the application. 
% To tackle ENO, EH solutions requires prediction of EH patterns which 
% Therefore, the ultimate enabler of the judicious usage of the available energy is an energy management algorithm. 
% For example, consider a setting where the harvested energy is below a certain level at a given period and the wearable device needs to send an urgent data to a local hub through a wireless communication protocol, which is a power-hungry operation. 
% A naive approach to remedy this is increasing the harvested energy and reducing the power consumption of the operations on the target device.
% Although tackling such issues are significant, the dynamic nature of the harvested energy suggests that the ultimate enabler of the judicious usage of the available energy is an energy management algorithm. 
These algorithms should satisfy the following conditions to be deployed on a resource-constrained device: \textit{(i)} incurring low execution time and power consumption overhead, \textit{(ii)} having a small memory footprint, \textit{(iii)} being responsive to the changes in the environment, and ideally, \textit{(iv)} learning to adopt such changes.
To this end, our goal is to develop a lightweight energy manager that enables ENO while maximizing the utilization of the device under dynamic energy constraints and EH conditions.

This paper presents a reinforcement learning (RL) based energy management framework, tinyMAN, for resource-constrained wearable edge devices.
%In a given time interval (e.g. within an hour), 
tinyMAN takes the battery level and the previous harvested energy values as inputs (states) and maximizes the utility of the device by judiciously allocating the harvested energy throughout the day (action). 
It employs Proximal Policy Optimization (PPO) algorithm, which is a state-of-the-art RL algorithm for continuous action spaces ~\cite{schulman2017proximal}.
Hence, the energy allocation values that tinyMAN yields can take continuous values according to the current energy availability.
Over time, by interacting with the environment, the agent learns to manage the harvested energy on the device according to battery energy level and the harvested energy. 
To achieve this, 
%To this end, 
we first develop an environment for the RL agent to interact with.
This environment makes use of the light and motion EH modalities and \textit{American Time Use Survey}~\cite{amtus} data from 4772 different users to model the dynamic changes in the harvested energy and battery. 
Then, we design a generalized reward function that defines the device utility as a function of the energy consumption.
The nature of the reward function also enables compatibility with any device and application.
% Additionally, the reward function includes two constraints on the battery: \textit{(i)} the minimum battery energy level at any given time and  \textit{(ii)} the target battery energy level at the end of the day, which should be equal to the initial battery at the start of the day.
% The first constraint maximizes the utility and ensures that the device does not drain the battery with excess amount of energy allocation.
% The second constraint ensures energy neutral operation (ENO) such that the device sustains itself without any charging requirement.

% the initial battery energy level at the start of the day, current battery energy level at the given hour of the day, the harvested energy from the previous hour and the cumulative harvested energy throughout the day, and lastly, the current hour in a day.

tinyMAN is trained on a cluster of users with randomly selected initial battery energy levels and EH conditions.
Therefore, it is responsive to various EH and battery energy level scenarios.
We compare our approach to prior approaches in the literature and also with an optimal solution. This comparison shows that tinyMAN achieves up to 45\% higher utility values.
Furthermore, we deploy our framework on a wearable device prototype to measure the execution time, energy consumption, and memory usage overhead.

\noindent \textit{The major contributions of this work are as follows:}
\vspace{-1mm}
\begin{itemize}[leftmargin=*]
    \item We present tinyMAN, a \emph{prediction-free} RL based energy manager for resource-constrained wearable edge IoT devices,  
    \item tinyMAN achieves 45\% higher device utilization than the state-of-the-art approaches by learning the underlying EH patterns for different users while maintaining energy neutrality,
    \item tinyMAN is easily deployable on wearable devices thanks to its small memory footprint of less than 100 KB and energy consumption of 27.75 $\upmu$J per inference.
\vspace{-1mm}
\end{itemize}
% (i) provides  a generalized reward function that supports any device and application, (ii) is a prediction-free energy management framework, and (iii) is TICC2652R system-on-chip for processing on the edge,

In the rest, 
Section \ref{sec:related_work} reviews the related work, 
while Section \ref{sec:Overview} introduces the problem formulation and describes the PPO algorithm.
Section \ref{sec:approach} formulates the environment dynamics and presents the proposed energy manager, tinyMAN. 
Finally, we evaluate and discuss the results in Section \ref{sec:evaluation} and conclude the paper in Section \ref{sec:conclusion}.

%\vspace{-1mm}
\section{Related Work} \label{sec:related_work}
% \vspace{-1mm}

% Prior work considers four main wearable EH modalities: light, motion, electromagnetic waves, and heat~\cite{sudevalayam2010energy}. 
% Among these, ambient light EH offers the highest capacity of over 1 mW outdoors (5000 lux) and close to 100 $\upmu$W indoors (500 lux) with an 8.1-cm$^2$ flexible PV-cell~\cite{jokic2017powering}.
% Human motion EH can harvest about 15 $\upmu$W with a 23.8-cm$^2$ piezoelectric patch while the wearer is walking~\cite{tuncel2020towards}.
% Similarly, radio-frequency (RF) EH can harvest 10 $\upmu$W with an 18.4-cm$^2$ flexible antenna with a signal strength of -10 dBm at 915 MHz~\cite{nguyen2018hybrid}, and 
% body-heat EH has power levels of about 3 $\upmu$W with a 1-cm$^2$ flexible harvester at an ambient temperature of 15$\degree$C (i.e., a temperature difference of 22$\degree$C)~\cite{huu2018flexible}.
% In this work, we use the combination of light and motion EH modalities.

%ENERGY NEUTRAL OP
Energy harvesting devices aim for ENO to achieve self-sustainability.
Kansal et al.~\cite{kansal2007power}, ensure
ENO if the total energy consumed in a given period is equal to the harvested energy in the same period.
The authors propose a linear programming approach to maximize the duty cycle of a sensor node and a lightweight heuristic to help solve the linear programming with ease.
Although their approach is lightweight, it does not consider the application requirements when deciding the duty cycle of the nodes.
Bhat et al. address this issue by using a generalized utility function that defines the application characteristics~\cite{bhat2017near}. 
They presented a lightweight framework based on the closed-form solution of the optimization problem that maximizes the utility while maintaining ENO.
However, the framework can yield sub-optimal solutions since the closed-form solution is obtained by relaxing one of the constraints in the original problem.
In addition, both approaches depend on a predictive model for the future EH values.
\textit{Thus, their performances are highly dependent upon the accuracy of the predictions}.

\begin{table}[b]
\vspace{-2mm}
\caption{Related work in energy management}
\label{tab:RWtable}
\vspace{-3mm}
\small
\begin{tabular}{@{}rccc@{}}
\toprule
\textbf{Ref} & \textbf{Generalized Reward} & \textbf{Prediction Free} & \textbf{Deployable} \\ \midrule
\cite{kansal2007power} & \ding{55} & \ding{55} & \ding{51} \\
\cite{bhat2017near} & \ding{51} & \ding{55} & \ding{51} \\
\cite{aoudia2018rlman} & \ding{55} & \ding{51} & \ding{55} \\
\rowcolor{lightgray}
tinyMAN & \ding{51} & \ding{51} & \ding{51} \\ \bottomrule
\end{tabular}
\end{table}

Prediction-free approaches do not rely on forecasts of the harvested energy, in contrast to the prediction-based approaches presented above~\cite{aoudia2018rlman}.
RLMan is a recent prediction-free energy management approach based on reinforcement learning~\cite{aoudia2018rlman}. 
It aims to maximize packet generation rate while avoiding power failures.
Although it shows significant improvements in average packet rate, the reward function in RLMan focuses on maximizing the packet rate in a point-to-point communication system, which does not generalize to other performance metrics and ignores application requirements.
In addition,  the authors do not discuss the deployability of their framework on edge devices.
In complement to the previous studies, we present tinyMAN, a prediction-free energy manager which uses a generalized reward function and is easily deployable on resource-constrained edge devices, as shown in Table~\ref{tab:RWtable}.
\textit{Furthermore, we provide open-source access to the trained models and to our codebase.}
% We evaluate \todo{tinyMAN} on a wearable device prototype and demonstrate its performance and deployability.

% \vspace{-6mm}
\section{Background}
\label{sec:Overview}
This section first introduces the battery energy dynamics and constraints to formulate the optimization problem. 
It also explains how various EH patterns are obtained. 
Then, it describes the Proximal Policy Optimization algorithm used to train the tinyMAN RL agent. 

\subsection{Problem Formulation}
\label{sec:dynamics}

% \todo{Explain the device, what it does, what components it contains etc.. \\
% Explain how the values/parameters are determined using this device,\\
% (use the table for ECO review),\\
% Explain how it behaves in the environment, battery dynamics (definitions of EBt EAt EHt etc..), constraints, utility, what are alpha, beta etc..\\
% Finally, how does the agent interact with the environment}

\begin{figure}[t]
	\centering
	\includegraphics[width=0.8\linewidth]{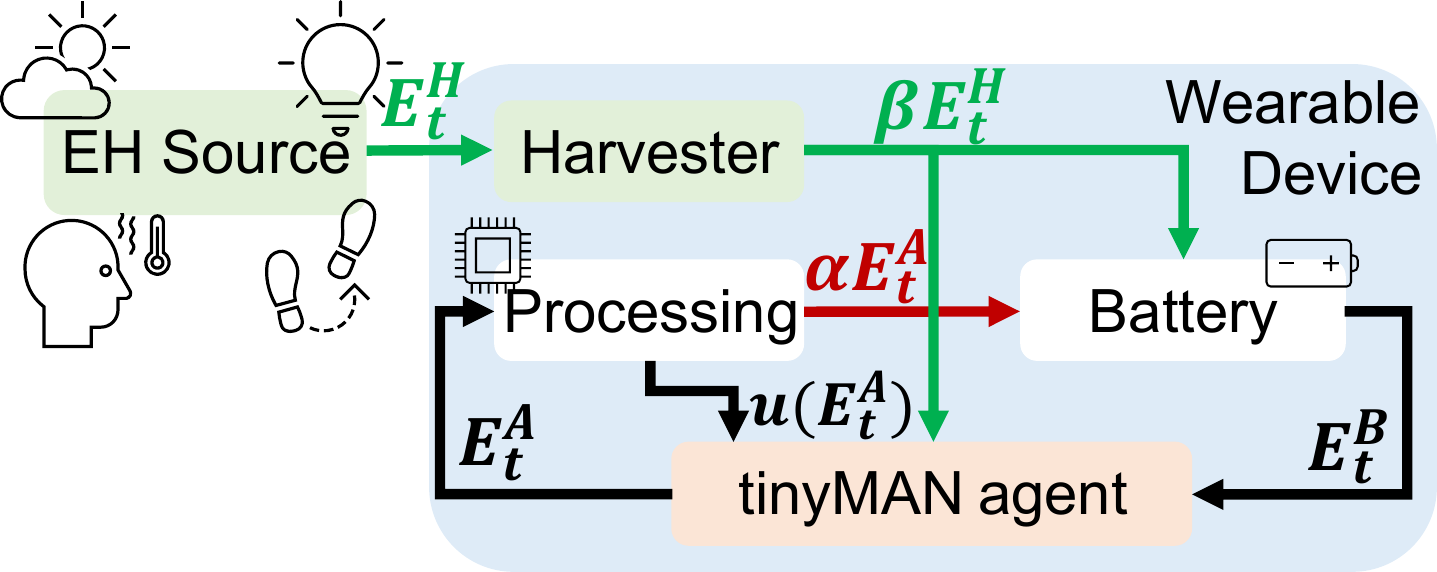}
	\vspace{-2mm}
	\caption{Illustration of the environment.}
	\label{fig:environment}
	\vspace{-5mm}
\end{figure}

The proposed tinyMAN framework is deployed in an environment that consists of a target device and an EH source, as depicted in Figure~\ref{fig:environment}. 
In the following, we define the battery energy dynamics, the relevant constraints, and the utility function of the device and explain the EH source model.
% Then, we present the optimization problem 
% Finally, we 

\noindent \textbf{Battery dynamics and constraints:}
tinyMAN finds the optimum energy allocations that maximize the utilization of a target device under ENO and battery constraints.
In this work, we use a prototype wearable device as the target platform to deploy tinyMAN.
The device houses a flexible, small form-factor LiPo battery with a capacity of 12 mAh, and can charge the battery through energy harvesting.
Therefore, the battery energy dynamics in the environment is a function of:
\begin{enumerate}
    \item $E^B_t$ the battery energy level at the start of time interval $t$
    \item $E^A_t$ the allocated energy at the start of time interval $t$
    \item $E^H_{t}$ the harvested energy in time interval $t$
\end{enumerate}
Our energy management framework uses an episodic setting where each episode corresponds to a single day ($T=$ 24 hours), and each step $t$ in an episode corresponds to an hour.

Using these definitions, we write the battery energy dynamics as follows:
\begin{equation}\label{eqn:battery_energy}
E^B_{t+1} = E^B_t + \beta E^H_{t} - \alpha E^A_t, \hspace{1mm} t \in T %\mathcal{S_T}
% \hspace{2mm} \mathrm{and} \hspace{2mm} E^B_{T} = E^B_0
\end{equation}
where $\beta$ corresponds to the efficiency of the harvester and $\alpha$ corresponds to the percent utilization of the allocated energy (i.e., $\alpha E^A_t$ is the actual consumed energy).

There are two physical constraints on the battery level. It is bounded from below at zero and from the top at the battery capacity ($E^B_{cap}$). Furthermore, we want the device to have an emergency reservoir at all times to serve as backup energy:
\begin{equation}\label{eqn:energyconstraint}
E^B_{cap} \geq  E^B_{t} \geq  E^B_{min}, \hspace{2mm}   t \in T %\mathcal{S_T}
\end{equation}
%

%
% \begin{equation}\label{eqn:eminconstraint}
% E^B_{t} \geq  E^B_{min} \hspace{2mm}   t \in T %\mathcal{S_T}
% \end{equation}
%

To achieve ENO, tinyMAN ensures that the battery energy level at the end of an episode is equal to a specified target:
\begin{equation} \label{eqn:targetconstraint}
% E^B_{T} \geq E_{target} \hspace{2mm} 
E^B_{T} \approx E_{target} \hspace{2mm} 
\end{equation}
For achieving ENO, we set $E_{target}=E^B_0$ such that the battery energy level at the end of the episode is equal to the battery energy level at the beginning of the same episode.
We enforce these constraints using the reward function as explained in Section~\ref{sec:env_dynamics}.
% Similar to the previous $E^B_{min}$ constraint, we also enforce the target energy constraint using the reward function.

\noindent \textbf{Device utility:} 
The utilization of the device is a metric that represents the useful output produced by the device, such as accuracy or throughput, depending on the target application running on the device.
For example, for human activity recognition, a state-of-the-art application that utilizes a low-power wearable device, the utility is defined by the classification accuracy.
Nonetheless, tinyMAN supports any arbitrary utility function.
% The most notable advantage of tinyMAN is that it supports any arbitrary utility function.}

For the current work, we define the utility according to the minimum energy consumption of the device in an hour.
Specifically, the device utility is zero (or negative) if the allocated energy is less than the minimum energy consumption of the device in a given interval.
We list the components used in the wearable device prototype in Table~\ref{tab:components} to calculate the minimum energy consumption in an hour.
% was developed for data collection for human activity recognition applications,
According to these values, the sum of the idle currents of the components amounts to 54.6 $\upmu$A, and the idle energy consumption of the device in an hour is $E^A_{min}=$ 0.64 J with a VDD of 3.3V.
Therefore, the device utility will vanish if $E^A_t < E^A_{min}$ (i.e., the device does not produce any useful output).
For $E^A_t > E^A_{min}$, \emph{the utility function can have any shape according to the needs of the application}. 
For this work, we used a logarithmic utility function with a diminishing rate of return, as elaborated in Section~\ref{sec:env_dynamics}.

% We stress that this logarithmic utility function is very general, which allows different target applications and devices to be captured by changing the $E^A_{min}$.

% Using this value, we define the utility of the device as a function of $E^A_t$ with a diminishing rate of return:
% %
% \begin{equation}\label{eqn:log_utility}
% u(E^A_t) = \ln\bigg(\frac{E^A_t}{E^A_{min}} \bigg)
% \end{equation}
% %
% We stress that this logarithmic utility function is very general, which allows different target applications and devices to be captured by changing the $E^A_{min}$.

\begin{table}[t]
\caption{Components used in the prototype wearable device.}
\vspace{-2mm}
\label{tab:components}
\footnotesize
\begin{tabular}{@{}lllll@{}}
\toprule
\textbf{Component} & \textbf{VDD} & \textbf{I$_{idle}$} & \textbf{I$_{active}$} & \textbf{Part \#} \\ \midrule
Microcontroller & 1.8-3.8V & 0.9 $\upmu$A & \begin{tabular}[c]{@{}l@{}}Sensor Cont.: 30 $\upmu$A\\ Active: 3.4 mA\end{tabular} & CC2652R \\
IMU & 1.7-3.6V & 8 $\upmu$A & \begin{tabular}[c]{@{}l@{}}Acc only: 450 $\upmu$A\\ Gyro only: 3.2   mA\end{tabular} & MPU9250 \\
Nonvolatile Ram & 1.6-3.6V & 10 $\upmu$A & \begin{tabular}[c]{@{}l@{}}Rewrite: 1.3 mA \\ Read-out: 0.2 mA\end{tabular} & MB85AS4MT \\
\begin{tabular}[c]{@{}l@{}}Humid. \& Temp. \\ Sensor\end{tabular} & 2.7-5.5V & 0.1 $\upmu$A & 1 Hz: 1.2 $\upmu$A & HDC1000 \\
\begin{tabular}[c]{@{}l@{}}Ambient Light \\ Sensor\end{tabular} & 1.6-3.6V & 0.3 $\upmu$A & 1.8 $\upmu$A & OPT3001 \\
\begin{tabular}[c]{@{}l@{}}Boost Converter \\ for EH\end{tabular} & 2.5-5.2V & 0.3 $\upmu$A & - & BQ25504 \\
LDO linear regulator & 2.0-5.5V & 35 $\upmu$A & - & TLV702 \\ \bottomrule
\end{tabular}
\vspace{-5mm}
\end{table}

\begin{figure*}[t]
\centering
\includegraphics[width=1\textwidth]{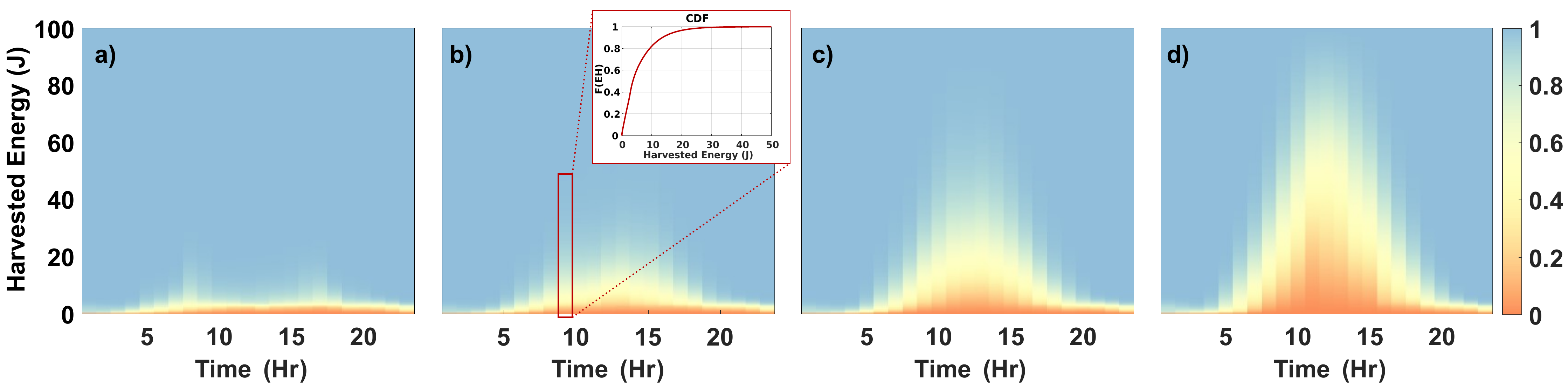} 
\vspace*{-6mm}
\caption{Cumulative distribution function of the harvested energy for a) Cluster 1, b) Cluster 2, c) Cluster 3, and d) Cluster 4}
\label{fig:ehmodel}
\normalsize
\end{figure*}

\noindent \textbf{EH Source:}
The EH source uses the dataset presented in~\cite{tuncel2021much} to generate EH scenarios according to different user patterns.
This dataset uses the combination of light and motion energy as the ambient energy sources, and it combines power measurement data with the activity and location information of 4772 users from the American Time Use Survey dataset~\cite{amtus} to generate varying 24-hour EH patterns per user.
We divide the EH dataset~\cite{tuncel2021much} into four clusters according to the users' EH patterns throughout the day.
The hourly distributions of these four clusters are illustrated in Figure~\ref{fig:ehmodel}.
These distributions are based on the mean and the standard deviation of EH patterns in the same cluster.
% , \todo{as illustrated in Figure~\ref{fig:ehmodel}.
% In this figure, the black dots show the mean expected energy harvesting for the corresponding hour, whereas the shaded bands represent the variance of the distribution for the same hour.
Therefore, the EH source generates a harvested energy value at every hour according to the distributions in the dataset as the day progresses.

\subsection{Proximal Policy Optimization}
\label{sec:ppo}

The main objective of an RL agent is to maximize the cumulative rewards by interacting with the environment.
According to the state $s$ of the environment and the current policy $\pi$, the agent chooses an action $a$.
Based on this action, environment returns next state $s^{\prime}$ and reward $r$.
Th environment is initialized with state $s_0$ (start of the day, $t=$ 0) and terminates after $T=$ 24 steps (end of the day, $t=$ 24).
The policy $\pi$ is represented by a neural network with parameters $\theta$.
The agent interacts with the environment using the current policy $\pi_\theta$ and collects samples ($s,a,r,s^\prime$).
In policy gradient algorithms, the policy network is updated using the gradient of the policy multiplied with discounted cumulative rewards as a loss function and plugging it into the gradient ascent algorithm.
This update is generally done using samples from multiple episodes.
The discounted cumulative rewards can exhibit high variations since each episode follows a different trajectory based on the actions.
To reduce this variance, a bias is introduced as an advantage function that measures the benefit of taking action at a given state.
The loss function then takes the form:
\begin{gather}  
 L_\theta = \sum_{n=0}^{N}\sum_{t=0}^{T}log\pi_\theta(a_t|s_t)A(s_t,a_t)  \label{eq:policy_loss} \\
 A(s_t,a_t) = r_t +\gamma V_\phi(s_{t+1})-V_\phi(s_{t}) \label{eq:adv} 
\end{gather}
Here, $\pi_\theta(a_t|s_t)$ is the current policy which gives the probability of taking action $a$ in state $s$.
Advantage function is represented by $A(s_t,a_t)$ and is given by Equation~\ref{eq:adv} where $\gamma~\in [0,1]$ is the discount factor and $V_\phi(s)$ is the value network which estimates the expected discounted sum of rewards for a given state $s$. $N$ is the number of episodes, and $T$ is the number of steps in an episode.
Value network $V_\phi(s)$ is also updated during training using gradient descent with the mean-squared error between target values and the estimated values as a loss function:
\begin{gather}  
L_\phi = \frac{1}{NT}\sum_{n=0}^{N}\sum_{t=0}^{T}(V_\phi(s_{t}) - (r_t +\gamma V_\phi(s_{t+1})))^2  \label{eq:value_loss}
\end{gather}

PPO aims at improving the training stability by avoiding network parameter updates that change the policy drastically at each step of optimization. To this end, it modifies the policy loss (Equation~\ref{eq:policy_loss}) in such a way that the distance between new policy ($\pi_\theta(a|s)$) and the old policy (${{}\pi_\theta}_{old}(a|s)$) is enforced to be small. It achieves its goal using the following loss function:
\begin{multline}
\label{eq:ppo_loss}
L_\theta^{PPO} = \frac{1}{NT}\sum_{n=0}^{N}\sum_{t=0}^{T}min(\rho(\theta)A_t,
~clip(\rho(\theta), 1-\epsilon, 1+\epsilon)A_t)  
\end{multline}
\begin{gather}
\rho(\theta) = \frac{\pi_\theta(a_t|s_t)}{{{}\pi_\theta}_{old}(a_t|s_t)} \label{eq:ratio_loss} 
\end{gather}
In this equation,  ${{}\pi_\theta}_{old}(a|s)$ is the policy that is used to collect samples by interacting with the environment and $\pi_\theta(a|s)$ is the policy that is being updated using this loss function.
PPO imposes a limitation on the distance between ${{}\pi_\theta}_{old}(a|s)$ and $\pi_\theta(a|s)$ by clipping the ratio $\rho(\theta)$ between two distribution with $\epsilon$ where $\epsilon$ is a hyperparameter of the algorithm.
An entropy term may also be included in this loss function to encourage sufficient exploration.
\section{Proposed Energy Manager -- \lowercase{tiny}MAN}
\label{sec:approach}

This section provides the environment dynamics and introduces the RL framework, the core algorithm used in tinyMAN.

\subsection{Environment Dynamics}
\label{sec:env_dynamics}
Our goal is to maximize the utilization of the device under certain battery energy level constraints.
In our framework, the environment dynamics are determined to enable the adaptability of tinyMAN by any device and application.

\noindent\textbf{State Space:} The state is a 5-tuple that consists of: 
% \vspace{-1.5mm}
\begin{itemize}[leftmargin=*]
    \item[-] \textbf{Current battery energy level ($E_t^B \in [0, E^B_{cap}]$):} The energy level of the battery at the beginning of the current step $t$. 
    % We clip this value on the higher end according to the physical characteristic of the battery to ensure that the battery is full.
    \item[-] \textbf{EH from the previous time step ($E^H_{t-1}\in \mathbb{R}$):} Harvested energy during the previous step $t-$1.
    \item[-] \textbf{Initial battery energy level ($E^B_{0}\in \mathbb{R})$:} The energy level of the battery at the beginning of the episode ($t$=0).
    \item[-] \textbf{Time ($t\in \mathbb{Z}$):} The current step $t$, which corresponds to the current hour of the day.
    \item[-] \textbf{Cumulative EH ($\sum_{\tau=0}^{t-1}E^H_{\tau}\in \mathbb{R}$):} Cumulative harvested energy in the previous time steps.
\end{itemize}

\noindent\textbf{Action Space:} The action is the allocated energy at every time step ($E_t^A\in [E_{min}^A, E_t^B]$). Since the application on the device needs a minimum energy level to stay in the idle state, we set a minimum level constraint on the action ($E_{min}^A$).

\noindent\textbf{Reward function:} Our objective is to maximize the utility of the device under certain constraints on the battery energy level. 
tinyMAN supports any arbitrary utility function, but to have a fair comparison with the literature~\cite{bhat2017near}, 
%  and to satisfy the requirements elaborated in Section~\ref{sec:dynamics},
we use the following logarithmic utility function in this work: 
\begin{equation}\label{eqn:log_utility}
u(E^A_t) = \ln\bigg(\frac{E^A_t}{E^A_{min}} \bigg)
\end{equation}
In an RL setting, the constraints on the battery can be imposed by the reward function. 
There are two constraints that can be imposed to the reward function: \textit{(i)} emergency reservoir energy constraint (Equation~\ref{eqn:energyconstraint}) and \textit{(ii)} ENO constraint (Equation~\ref{eqn:targetconstraint}).
Considering the objective and the constraints on the battery, the reward function becomes:
\begin{equation}
r_t =\begin{cases} 
      u(E^A_t) & E_t^B\geq E_{min}^B~and~t\neq T\\
      u(E^A_t)  - (E_{min}^B - E_t^B)^2 & E_t^B\le E_{min}^B~and~t\neq T  \\
      -(E_t^B - E_{target})^2 & t=T 
   \end{cases}
\end{equation}
Here, we impose the emergency reservoir energy constraint using the term $-(E_{min}^B - E_t^B)^2$ and the ENO constraint using the term $-(E_t^B - E_{target})^2$. 
Moreover, an episode terminates if time $T$ is reached or the battery is completely drained.

According to the environment dynamics explained in this section, we develop our environment in Python and register it as an OpenAI's Gym~\cite{gym} environment. 

% \vspace{-3mm}
\begin{algorithm}[b] 
\small
\caption{tinyMAN - RL based Energy Manager } \label{algo:ppo_algo}
% \SetAlgoLined
\SetAlgoVlined
\SetNoFillComment
% \begin{algorithmic}
Initialize policy and value network with parameters $\theta_0$ and $\phi_0$\\
% Initialize value network with parameters and  $\phi_0$ \\
Initialize random policy $\pi_{\theta_0}$, empty trajectory buffer $\mathcal{D}$ with size $\mathbb{D}$ \\
% Initialize empty trajectory buffer $\mathcal{D}$ with size $\mathbb{D}$ \\
\For{n = \normalfont{0: $N$}} 
{
Initialize environment with randomly chosen initial\\
battery energy $E_0^B$ and EH patterns\\
\While {$\mathcal{D}$ is not full}{
\For{t = \normalfont{0: $T$}} {
Choose $a_t$ according the current policy $\pi_{\theta_i}$\\
Collect samples $\{s_t,a_t,r_t,s_t^\prime\}$ by interacting with \\
the environment using action $a_t$\\
}
}
Obtain $A_t$, $r_t + V_{\phi_i}(s_{t+1})$ and $\pi_{\theta_i}(a_t|s_t)$  \\
using policy and value networks (see Section \ref{sec:ppo} for details)\\
\For{ k = \normalfont{1: $K$}} {
\For{ b = \normalfont{0: $(\mathbb{D}/d)$}} {
$batch_{start} = d\times(b -1)$\\
$batch_{end} = d\times(b)$\\
$minibatch \leftarrow ~\mathcal{D} [batch_{start}:batch_{end}]$\\
% Obtain value estimates and new logarithm \\
% of probabilities to calculate the loss functions \\
$L \leftarrow -L_{\theta}^{PPO} + c_1L_{\phi} + c_2H(\pi_{\theta})$\\
Minimize the total loss L
}
}
$\theta_{i+1} \leftarrow ~ L_{\theta_K}^{PPO}$\\
$\phi_{i+1} \leftarrow ~ L_{\phi_K}$\\
% \underset{\phi}{\arg\min} 
Clear $\mathcal{D}$
}
% \end{algorithmic}
\end{algorithm}
% \vspace{-3mm}
\subsection{Proposed RL Framework}
\label{sec:RL_framework}

Since the proposed tinyMAN framework is deployed on a wearable device, we first identify the characteristics of the target device such as battery capacity, minimum battery energy level ($E_{min}^B$), and minimum energy allocation ($E_{min}^A$). 
These characteristics do not change over time during the training. 
The EH dataset~\cite{tuncel2021much} is divided into four clusters according to the users' EH patterns throughout the day.
% the users which have significantly low EH patterns are discarded and 
% and use EH patterns from the same cluster during training. 
The agent is trained separately on each cluster.
% tinyMAN uses an episodic setting where each episode ($n$) corresponds to a single day and each time step ($t$) in an episode corresponds to an hour.
Specifically, at the beginning of each episode $n$, we randomly choose an initial battery energy level.
Then, we generate an EH pattern from the hourly distributions illustrated in Figure~\ref{fig:ehmodel}.
% These distributions are based on the mean and the standard deviation of EH patterns in the same cluster for that episode.
The generated EH pattern is different for each episode.
Thus, tinyMAN inherently learns the EH patterns of the users in that cluster.
The initial conditions and the EH patterns can differ significantly between different episodes. 
This may result in a high gradient variance and unstable learning progress during the training.
For this reason we employ PPO in our work, as it guarantees that policy updates do not deviate largely.
In addition, PPO uses little space in the memory, which fits the resource-constrained nature of the target device.
% This improves the generalizability of our approach.

Algorithm~\ref{algo:ppo_algo} describes the training of tinyMAN agent for a given cluster of users.
The agent starts the first episode with a random policy $\pi_{\theta_0}$ with parameters $\theta_0$.
Using the current policy $\pi_{\theta_i}$, the agent first collects samples until the trajectory buffer $\mathcal{D}$ with a predefined size of $\mathbb{D}$ is full.
Note that this trajectory buffer is not the experience replay buffer commonly used in off-policy RL algorithms.
Using the samples in the trajectory buffer, advantages $A_t$, target values $r_t + V_{\phi_i}(s_{t+1})$, and the probabilities $\pi_{\theta_i}(a_t|s_t)$ are obtained using the policy network $\pi_{\theta_i}$ and the value network $V_{\phi_i}$.
The algorithm updates both the policy and the value network parameters ($\theta$, $\phi$) according to the loss functions described in Section~\ref{sec:ppo}.
We augment the loss function for different networks and add an entropy term $H(\pi_{\theta})$ to increase the exploration of the algorithm. 
PPO updates the network parameters by generally taking multiple steps on minibatches.
The number of optimization steps $K$ and the minibatch size $d$, and the clipping value $\epsilon$ in the policy loss function are hyperparameters of the network. 
Both networks consist of fully connected layers with hyperbolic tangent as activation function. 
Additionally, the policy network also has a Gaussian distribution head to yield continuous values from a distribution. 
The number of hidden layers ($N_{Layer}$) and neurons ($N_{Neuron}$ ) are the same for both networks.

We implement tinyMAN in Python by utilizing PFRL~\cite{pfrl} library for the PPO algorithm using Adam optimizer with a learning rate of 1E-4. %$0.0001$.
The hyperparameters for tinyMAN are given in Table~\ref{tab:hyperparameters}.

\begin{table}[b]
\vspace{-4mm}
\small
\caption{Definition of the hyperparameters and their values.}
\vspace{-2mm}
\label{tab:hyperparameters}
\begin{tabular}{@{}lll@{}}
\toprule
\textbf{Hyperparameter} & \textbf{Description}   & \textbf{Value} \\ \midrule
$\alpha$                & Percent utilization       & 1           \\
$\beta$                 & Efficiency of the harvester        & 1          \\
$\gamma$                & Discount factor        & 1           \\
$N$                     & Number of episodes     & 200000         \\
$T$                     & Number of time steps   & 24             \\
$c_1$                   & Value loss coefficient & 0.5            \\
$c_2$                   & Entropy coefficient    & 0.01           \\
$\epsilon$              & Clipping factor        & 0.3            \\
$K$                     & PPO optimization steps             & 10             \\
$d$                     & Minibatch size         & 64             \\
$\mathbb{D}$                  & Trajectory buffer size & 2048           \\
$N_{Layer}$               & Number of hidden layers& 1           \\
$N_{Neuron}$               & Number of hidden neurons& \{16,32,64\}           \\
\bottomrule
\end{tabular}%
\vspace{-4mm}
\end{table}

\section{Experimental Evaluations} \label{sec:evaluation}
This section evaluates the tinyMAN framework from three aspects: (i) it presents the evolution of the tinyMAN agent during training, (ii) it compares the performance of the tinyMAN framework to two prediction-based prior approaches~\cite{kansal2007power, bhat2017near} in the literature, and (iii) it provides execution time, energy overhead and memory footprint measurements of the tinyMAN framework when deployed on a wearable device prototype.

% Four clusters (C1, C2, C3, and C4) are obtained from the dataset based on users' EH patterns.
% This section summarizes the performance of the four agents that are trained on these clusters.

\subsection{Training Evolution} 
\label{sec:train_evolve}

We first evaluate our agent's performance during training to highlight the evolution of a generalizable agent.
The harvested energy levels of the users are the lowest in cluster 1, and the highest in cluster 4, as depicted in Figure~\ref{fig:ehmodel}.
This section illustrates the results for cluster 2 since the users in this cluster are representative of an average person with low to intermediate levels of harvested energy during the day. 
Other clusters produce similar results.
Furthermore, we set the emergency reservoir energy as $E_{min}^B = 10$ J, which roughly corresponds to 5 minutes of active time for the components listed in Table~\ref{tab:components}.
We stress that this parameter can be tailored according to the requirements of another device or application.

% During the training, we randomly choose an initial battery energy level at the beginning of each episode.
%which satisfies the physical constraints of the battery.
% We also generate an EH pattern from a distribution that is based on the mean and the standard deviation of EH patterns in the same cluster for that episode.
% We also generate an EH pattern for the current episode from the hourly distributions shown in Figure~\ref{fig:ehmodel}.

\begin{figure*}[t]
\vspace*{-3mm}
% \small
\centering
\includegraphics[width=1\textwidth]{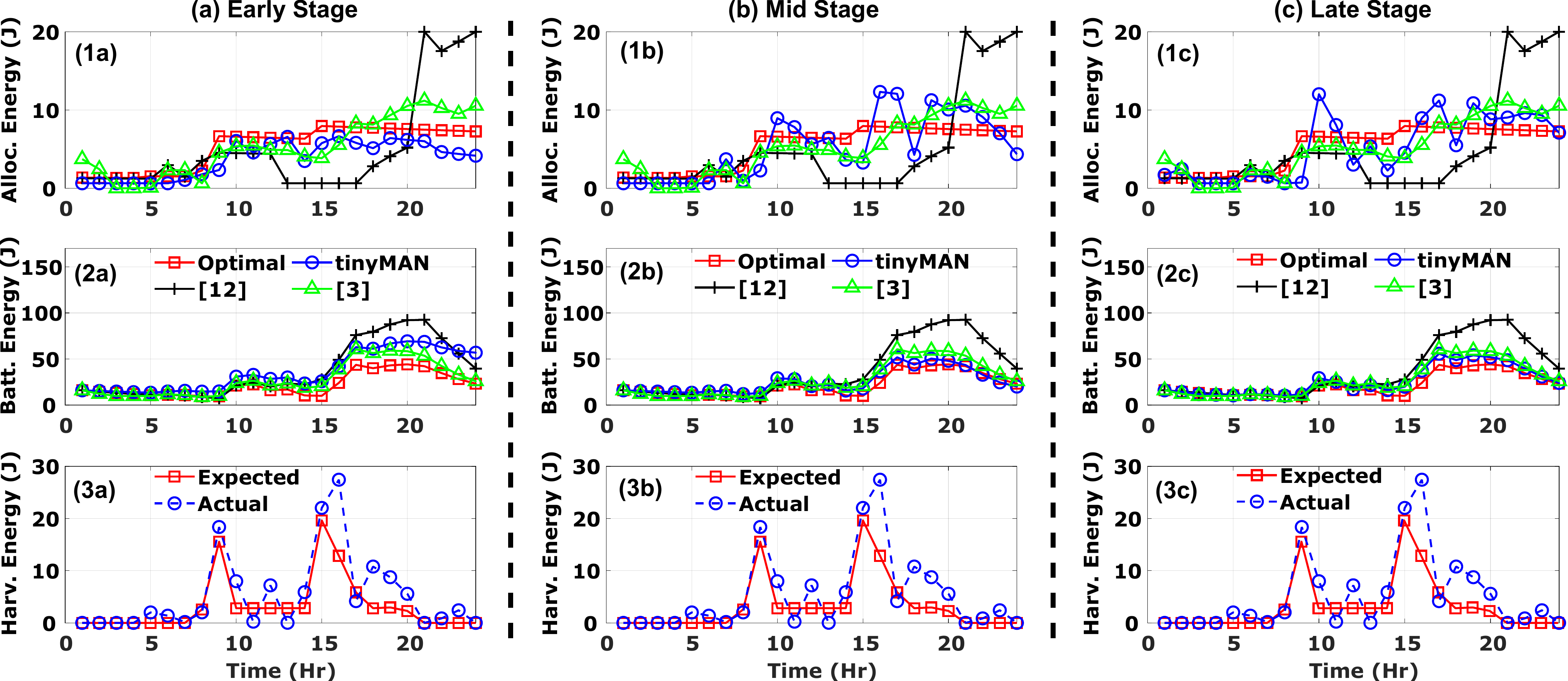} 
% \vspace*{-4mm}
\caption{Policy followed by tinyMAN agent at different stages of training for the median user of cluster 2. ($N_{Neuron}=$ 64).}

% \vspace*{-3mm}
\label{fig:trainvis}
\normalsize
% \vspace*{-3mm}
\end{figure*}

%We evaluate the training progress 
Figure~\ref{fig:trainvis} shows the allocated energy, battery energy level, and the expected/actual EH patterns for the median user in cluster 2 during training.
We follow the training steps described in Section~\ref{sec:RL_framework}.
The initial battery energy level $E^B_{0}$ is set as 16 J, which corresponds to 10\% of the battery.
% These distributions are based on the mean and the standard deviation of EH patterns in the same cluster for that episode.
% We also evaluate the training progress for the median user considering the total harvested energy levels in \todo{cluster 2}.
At the early stages of the training, tinyMAN takes conservative actions as shown in Figure~\ref{fig:trainvis} (1a). 
This suggests that the target energy level constraint (i.e., $E_T^B > E_{target}$) penalty is dominating the agent in these early stages.
% \todo{Normalized utility value for tinyMAN with respect to oracle is \todo{$XX$} in this stage. }
As the training progresses, actions that the agent takes are in correlation with the harvested energy since tinyMAN starts to learn a generalized representation of the EH patterns in this cluster.
Specifically, energy allocations increase as the EH increases and decrease as the EH decreases. 
This behavior and the fact that the constraints are satisfied can be seen in Figure~\ref{fig:trainvis} (b) and (c). 
% tinyMAN achieves \todo{$XX$} and \todo{$XX$} normalized utility at the mid and end stage if the training respectively while satisfying the energy constraints explained in Section~\ref{sec:dynamics}.

In addition to the behavior of the tinyMAN agent, we also illustrate the energy allocations computed by two prior prediction-based approaches in the literature~\cite{bhat2017near, kansal2007power}.
As both of these approaches are prediction-based, they use the specific expected EH pattern for a user, depicted with the red line in Figure~\ref{fig:trainvis} (3a, 3b, 3c).
On the contrary, tinyMAN implicitly learns the actual EH patterns during training, making it a prediction-free approach.
Finally, we compare our results against the optimal solution obtained by an offline solver (e.g., CVX) using the actual harvested energy during the day.
Although this solution is unfair and unrealistic, it provides an anchor point for assessing the quality of the energy allocations.
It can be seen that tinyMAN's actions oscillate around the optimal values with the red line in Figure~\ref{fig:trainvis} (1b, 1c).
%as it provides the maximum theoretical utility achievable by the given harvested energy.
%and gives decisions according to the current state

\subsection{Performance Evaluation}

We evaluate the performance of tinyMAN with three model sizes: $N_{Neuron}=$ \{16, 32, 64\}.
Similar to Section~\ref{sec:train_evolve}, we compare the performance of tinyMAN to two prior prediction-based approaches in the literature~\cite{bhat2017near, kansal2007power}, and also to an optimal solution. 
% obtained by an offline solver using the actual harvested energy during the day.
% We emphasize that this optimal solution is unfair and unrealistic, but it serves as a reference point as it provides the maximum theoretical utility achievable by the given harvested energy.
For a fair evaluation, we exclude randomly selected 10\% of the users in a cluster during training.
Then, using the energy harvesting patterns of these users, we compute the total utility obtained at the end of the day as follows:
\vspace{-3mm}
\begin{equation}\label{eqn:sum_utility}
U = \sum^T_{t=0} u(E^A_t) 
\end{equation}
For each cluster and tinyMAN model size, we evaluate the performance of our approach at four different initial battery energy levels: $E^B_0 = \{16, 48, 112, 144\}~J$. 
Table~\ref{tab:utility} presents the average total utility obtained from these four conditions for all approaches.
For a model size of 64, tinyMAN achieves up to 45\% and 10\% higher utility values than \cite{kansal2007power} and \cite{bhat2017near} while staying within at least 83\% of the optimal utility. 
Similarly, tinyMAN achieves up to 44\% higher utility values compared to prior approaches.
The utility achieved by tinyMAN decreases with smaller model sizes. This behavior is expected as the information captured by the network degrades.
% Similarly, tinyMAN achieves up to 44\% and 44\% higher utility values than \cite{kansal2007power} for model size of 32 and 16 respectively.
% However, compared to \cite{bhat2017near}, the increase in utility drops down to 5\% and for some clusters \cite{bhat2017near} has higher utility values than tinyMAN (e.g. for cluster 3) for model size of 32 and 16.
% Since the information captured by the network degrades as the model size of RL algorithm decreases, the observed decrease in utility is expected.
Moreover, we observe that for all solutions, in general, as the harvested energy increases from cluster 1 to cluster 4, the total utility increases since the available energy to allocate on the device increases. 
% However, there is only single out of common case where tinyMAN with model size of 16 obtains lower total utility for cluster 3 than for cluster 2. 
We emphasize that tinyMAN is trained for various battery energy levels and EH patterns which are generated using only the cluster's EH distribution.
This and the performance evaluation support that it can easily adapt to unseen user-specific EH patterns and battery energy levels, making it a preferred energy manager on an edge device with uncertainties in harvested energy.

\begin{table}[b]
\vspace{-3mm}
\caption{Comparison of the average daily utility obtained by tinyMAN with different model sizes to other approaches.}
\vspace{-2mm}
\label{tab:utility}
\begin{tabular}{@{}lcccccc@{}}
\toprule
 & \multirow{2}{*}{\textbf{Optimal}} & \multirow{2}{*}{\textbf{\cite{bhat2017near}}} & \multirow{2}{*}{\textbf{\cite{kansal2007power}}} & \multicolumn{3}{c}{\textbf{tinyMAN}} \\ \cmidrule(l){5-7} 
 &  &  &  & \textbf{16} & \textbf{32} & \textbf{64} \\ \midrule
Cluster 1 & 29.5 & 25.5 & 18.7 & 25.1 & 26.4 & 25.5 \\
Cluster 2 & 42.0 & 35.3 & 26.3 & 37.9 & 37.9 & 38.1 \\
Cluster 3 & 52.4 & 43.1 & 34.5 & 35.9 & 41.8 & 44.7 \\
Cluster 4 & 61.5 & 46.5 & 41.9 & 46.4 & 50.1 & 51.2 \\ \midrule
Cluster Avg. & 46.4 & 37.6 & 30.3 & 36.4 & 38.8 & 39.9 \\ \bottomrule
\end{tabular}
% \vspace{-4mm}
\end{table}	

\subsection{Deployability} %(exec time, energy overhead, area overhead, utility at different model sizes/quantizations?)

\begin{figure}[t]
\centering
\vspace{-3mm}
\includegraphics[width=0.4\textwidth]{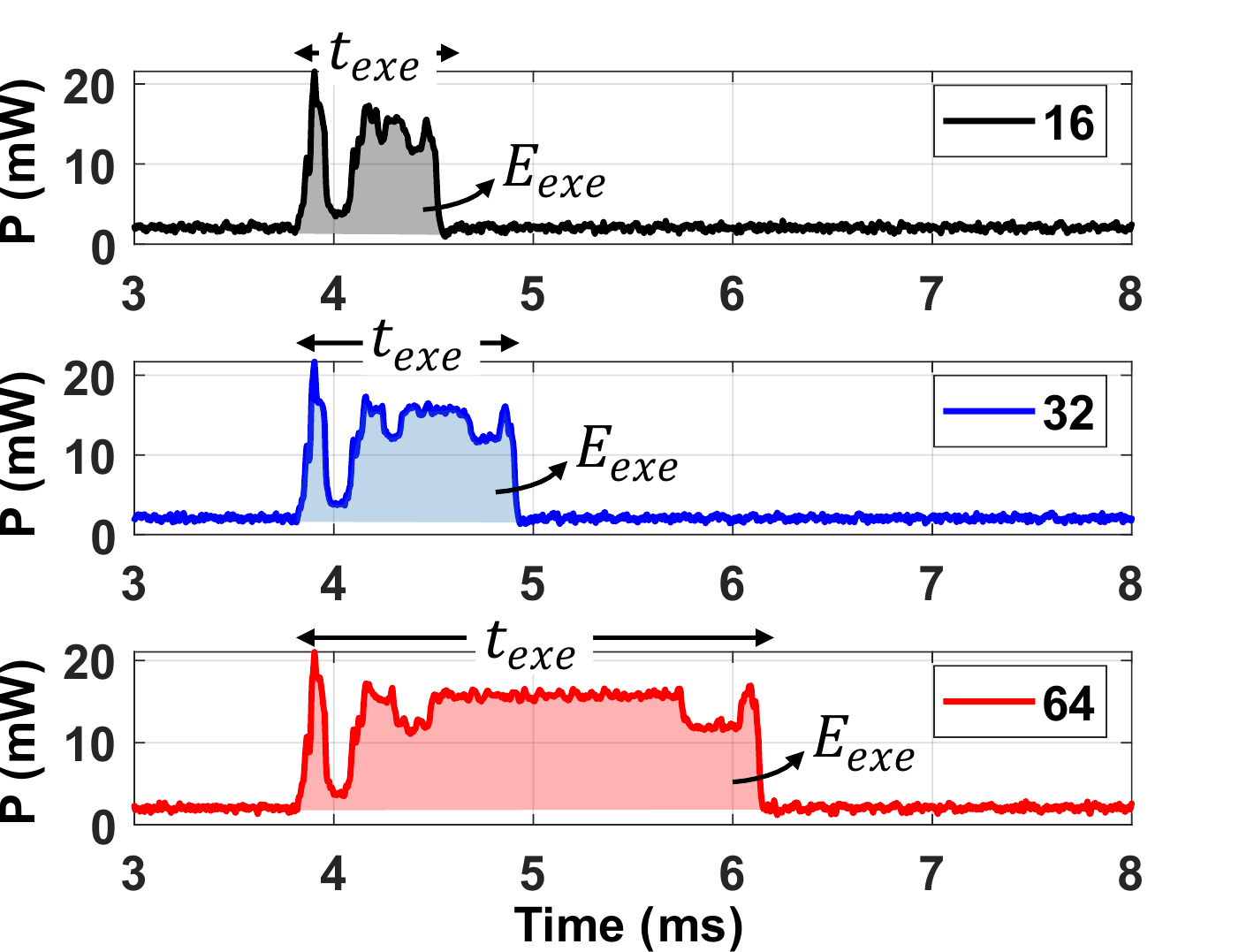} 
\vspace{-3mm}
\caption{Execution time and energy measurements of tinyMAN with different model sizes.}
\vspace{-5mm}
\label{fig:deployability2}
\end{figure}

The TI CC2652R microcontroller used on our prototype device incorporates an ARM Cortex M4F running at 48 MHz and has 352KB of flash memory and 80KB of SRAM.
These scarce resources highlight the importance of evaluating the trained models regarding their deployability on the target platform.
Therefore, we evaluate the deployability of the trained models from three aspects: (i) The execution time per inference, (ii) the energy consumption per inference, and (iii) memory utilization of the target hardware platform.
To do this analysis, we follow the Tensorflow Lite Micro (TFLM) flow to convert and deploy the trained models on the target device~\cite{david2020tensorflow}.
Then, we measure the current consumption of the TI microcontroller, as shown in Figure~\ref{fig:deployability2}.
Using these measurements, we calculate the execution time ($t_{exe}$) and energy consumption ($E_{exe}$) for different network sizes.
Finally, we use the ``Memory Allocation'' report of TI Code Composer Studio to obtain the memory utilization of the device.
Table~\ref{tab:deployability} summarizes our results.
The reported memory footprint is for the entire application, including necessary drivers and I/Os for debugging, such as UART and timers.
We also provide the utility values averaged over all clusters normalized with the optimal utility.
The device's execution time, energy consumption, and memory utilization decrease as the model size decreases.
% Likewise, the normalized utility also decreases.
Specifically, for a model size of 64, tinyMAN has a memory footprint of 91 KB and it consumes 27.75 $\upmu$J per inference.
% it achieves 0.86 normalized utility with 27.75 $\upmu$J energy consumption per inference.
When model sizes of 32 and 16 are used, tinyMAN's memory footprint reduces to 78 KB and 74 KB, respectively.
In addition, the energy consumption also reduces to  11.66 $\upmu$J and 6.74 $\upmu$J.
However, these reductions come at the expense of lower normalized utility.
Specifically, as model size decreases from 64 to 16, there is a 7\% reduction in the normalized utility.
% execution time, energy consumption, and the memory utilization of  becomes 2.36 ms,  and  respectively where it 
% Similarly, these values are 1.12 ms, 11.66 $\upmu$J, 78 KB and 0.75 ms, 6.75 $\upmu$J, 74 KB for model sizes 32 and 16 respectively.
In any case, these results suggest that tinyMAN is easily deployable on a resource-constrained wearable IoT device.

\begin{table}[b]
\vspace{-3mm}
\caption{tinyMAN's overhead for different model sizes.}
\vspace{-2mm}
\label{tab:deployability}
\begin{tabular}{@{}lllcl@{}}
\toprule
 & \textbf{\begin{tabular}[c]{@{}l@{}}Exec. \\ Time\end{tabular}} & \textbf{Energy} & \multicolumn{1}{c}{\textbf{\begin{tabular}[c]{@{}c@{}}Memory\\ {\small(Flash+SRAM)}\end{tabular}}} & \textbf{\begin{tabular}[c]{@{}l@{}}Norm. \\ Utility$^*$\end{tabular}} \\ \midrule
$N_{Neuron}$ = 16  & 0.75 ms & 6.75 $\upmu$J & 69KB+5KB &  0.79\\
$N_{Neuron}$ = 32  & 1.12 ms & 11.66 $\upmu$J & 73KB+5KB &  0.84\\
$N_{Neuron}$ = 64  & 2.36 ms & 27.75 $\upmu$J & 86KB+5KB &  0.86\\ \bottomrule
\end{tabular}
\\
\raggedright
\hspace{3mm} \textbf{$^*$}{\small The utility is normalized with respect to the optimal utility.}
\vspace{-3mm}
\end{table}

\section{Conclusion}
\label{sec:conclusion}

EH from ambient sources is an emerging solution to power low-energy wearable devices.
The harvested energy should be managed optimally to achieve energy-neutral operation and eliminate recharging requirements.
To this end, this paper presented tinyMAN, an RL-based prediction-free energy manager for resource-constrained wearable IoT devices.
tinyMAN judiciously uses the available energy to maximize the application performance while minimizing manual recharge interventions.
It maximizes the device utilization under dynamic energy harvesting patterns and battery constraints.
Additionally, tinyMAN is easily deployable on wearable IoT devices thanks to its small memory footprint being less than $100$ KB.
tinyMAN achieves up to 45\% higher device utilization than the prior approaches in the literature by inherently learning the EH patterns of users while consuming less than 27.75 $\upmu$J energy per inference.
As future work, we plan to extend our prototype device to log the harvested energy over a day. 
This will pave the way for adding online learning functionality to tinyMAN.
% , using the models in this work 


\begin{acks}
This work was supported in part by NSF CAREER award CNS-1651624, and DARPA Young Faculty Award (YFA) Grant D14AP00068.
\end{acks}

\bibliographystyle{ACM-Reference-Format}
{\vspace{0mm}\footnotesize{\bibliography{references/hdc, references/eh, references/rl}}}


\begin{thebibliography}{20}

%%% the Web address.

\ifx \showCODEN    \undefined \def \showCODEN     #1{\unskip}     \fi
\ifx \showDOI      \undefined \def \showDOI       #1{#1}\fi
\ifx \showISBNx    \undefined \def \showISBNx     #1{\unskip}     \fi
\ifx \showISBNxiii \undefined \def \showISBNxiii  #1{\unskip}     \fi
\ifx \showISSN     \undefined \def \showISSN      #1{\unskip}     \fi
\ifx \showLCCN     \undefined \def \showLCCN      #1{\unskip}     \fi
\ifx \shownote     \undefined \def \shownote      #1{#1}          \fi
\ifx \showarticletitle \undefined \def \showarticletitle #1{#1}   \fi
\ifx \showURL      \undefined \def \showURL       {\relax}        \fi
% invisible to TeX
\providecommand\bibfield[2]{#2}
\providecommand\bibinfo[2]{#2}
\providecommand\natexlab[1]{#1}
\providecommand\showeprint[2][]{arXiv:#2}

\bibitem[\protect\citeauthoryear{Aoudia, Gautier, and Berder}{Aoudia
  et~al\mbox{.}}{2018}]%
        {aoudia2018rlman}
\bibfield{author}{\bibinfo{person}{Fay{\c{c}}al~Ait Aoudia},
  \bibinfo{person}{Matthieu Gautier}, {and} \bibinfo{person}{Olivier Berder}.}
  \bibinfo{year}{2018}\natexlab{}.
\newblock \showarticletitle{RLMan: An energy manager based on reinforcement
  learning for energy harvesting wireless sensor networks}.
\newblock \bibinfo{journal}{\emph{IEEE Transactions on Green Communications and
  Networking}} \bibinfo{volume}{2}, \bibinfo{number}{2} (\bibinfo{year}{2018}),
  \bibinfo{pages}{408--417}.
\newblock


\bibitem[\protect\citeauthoryear{Basaklar, Tuncel, Narayana, Gumussoy, and
  Ogras}{Basaklar et~al\mbox{.}}{2021}]%
        {basaklar2021hypervector}
\bibfield{author}{\bibinfo{person}{Toygun Basaklar}, \bibinfo{person}{Yigit
  Tuncel}, \bibinfo{person}{Shruti~Yadav Narayana}, \bibinfo{person}{Suat
  Gumussoy}, {and} \bibinfo{person}{Umit~Y Ogras}.}
  \bibinfo{year}{2021}\natexlab{}.
\newblock \showarticletitle{Hypervector Design for Efficient Hyperdimensional
  Computing on Edge Devices}.
\newblock \bibinfo{journal}{\emph{arXiv preprint arXiv:2103.06709}}
  (\bibinfo{year}{2021}).
\newblock


\bibitem[\protect\citeauthoryear{Bhat, Park, and Ogras}{Bhat
  et~al\mbox{.}}{2017}]%
        {bhat2017near}
\bibfield{author}{\bibinfo{person}{Ganapati Bhat}, \bibinfo{person}{Jaehyun
  Park}, {and} \bibinfo{person}{Umit~Y Ogras}.}
  \bibinfo{year}{2017}\natexlab{}.
\newblock \showarticletitle{Near-optimal energy allocation for self-powered
  wearable systems}. In \bibinfo{booktitle}{\emph{IEEE/ACM International
  Conference on Computer-Aided Design}}. \bibinfo{pages}{368--375}.
\newblock


\bibitem[\protect\citeauthoryear{Bianchi, Bassoli, Lombardo, Fornacciari,
  Mordonini, and De~Munari}{Bianchi et~al\mbox{.}}{2019}]%
        {bianchi2019iot}
\bibfield{author}{\bibinfo{person}{Valentina Bianchi}, \bibinfo{person}{Marco
  Bassoli}, \bibinfo{person}{Gianfranco Lombardo}, \bibinfo{person}{Paolo
  Fornacciari}, \bibinfo{person}{Monica Mordonini}, {and}
  \bibinfo{person}{Ilaria De~Munari}.} \bibinfo{year}{2019}\natexlab{}.
\newblock \showarticletitle{IoT wearable sensor and deep learning: An
  integrated approach for personalized human activity recognition in a smart
  home environment}.
\newblock \bibinfo{journal}{\emph{IEEE Internet of Things Journal}}
  \bibinfo{volume}{6}, \bibinfo{number}{5} (\bibinfo{year}{2019}),
  \bibinfo{pages}{8553--8562}.
\newblock


\bibitem[\protect\citeauthoryear{Brockman, Cheung, Pettersson, Schneider,
  Schulman, Tang, and Zaremba}{Brockman et~al\mbox{.}}{2016}]%
        {gym}
\bibfield{author}{\bibinfo{person}{Greg Brockman}, \bibinfo{person}{Vicki
  Cheung}, \bibinfo{person}{Ludwig Pettersson}, \bibinfo{person}{Jonas
  Schneider}, \bibinfo{person}{John Schulman}, \bibinfo{person}{Jie Tang},
  {and} \bibinfo{person}{Wojciech Zaremba}.} \bibinfo{year}{2016}\natexlab{}.
\newblock \bibinfo{title}{OpenAI Gym}.
\newblock
\newblock
\showeprint{arXiv:1606.01540}


\bibitem[\protect\citeauthoryear{Capra, Peloso, Masera, Ruo~Roch, and
  Martina}{Capra et~al\mbox{.}}{2019}]%
        {capra2019edge}
\bibfield{author}{\bibinfo{person}{Maurizio Capra}, \bibinfo{person}{Riccardo
  Peloso}, \bibinfo{person}{Guido Masera}, \bibinfo{person}{Massimo Ruo~Roch},
  {and} \bibinfo{person}{Maurizio Martina}.} \bibinfo{year}{2019}\natexlab{}.
\newblock \showarticletitle{Edge computing: A survey on the hardware
  requirements in the internet of things world}.
\newblock \bibinfo{journal}{\emph{Future Internet}} \bibinfo{volume}{11},
  \bibinfo{number}{4} (\bibinfo{year}{2019}), \bibinfo{pages}{100}.
\newblock


\bibitem[\protect\citeauthoryear{David, Duke, Jain, Reddi, Jeffries, Li,
  Kreeger, Nappier, Natraj, Regev, et~al\mbox{.}}{David et~al\mbox{.}}{2020}]%
        {david2020tensorflow}
\bibfield{author}{\bibinfo{person}{Robert David}, \bibinfo{person}{Jared Duke},
  \bibinfo{person}{Advait Jain}, \bibinfo{person}{Vijay~Janapa Reddi},
  \bibinfo{person}{Nat Jeffries}, \bibinfo{person}{Jian Li},
  \bibinfo{person}{Nick Kreeger}, \bibinfo{person}{Ian Nappier},
  \bibinfo{person}{Meghna Natraj}, \bibinfo{person}{Shlomi Regev},
  {et~al\mbox{.}}} \bibinfo{year}{2020}\natexlab{}.
\newblock \showarticletitle{Tensorflow lite micro: Embedded machine learning on
  tinyml systems}.
\newblock \bibinfo{journal}{\emph{arXiv preprint arXiv:2010.08678}}
  (\bibinfo{year}{2020}).
\newblock


\bibitem[\protect\citeauthoryear{de~Lima et~al\mbox{.}}{de~Lima
  et~al\mbox{.}}{2017}]%
        {de2017feasibility}
\bibfield{author}{\bibinfo{person}{Ana L{\'\i}gia~Silva de Lima}
  {et~al\mbox{.}}} \bibinfo{year}{2017}\natexlab{}.
\newblock \showarticletitle{{Feasibility of Large-Scale Deployment of Multiple
  Wearable Sensors in Parkinson's Disease}}.
\newblock \bibinfo{journal}{\emph{PLOS One}} \bibinfo{volume}{12},
  \bibinfo{number}{12} (\bibinfo{year}{2017}), \bibinfo{pages}{e0189161}.
\newblock


\bibitem[\protect\citeauthoryear{Fraternali, Balaji, Sengupta, Hong, and
  Gupta}{Fraternali et~al\mbox{.}}{2020}]%
        {fraternali2020ember}
\bibfield{author}{\bibinfo{person}{Francesco Fraternali},
  \bibinfo{person}{Bharathan Balaji}, \bibinfo{person}{Dhiman Sengupta},
  \bibinfo{person}{Dezhi Hong}, {and} \bibinfo{person}{Rajesh~K Gupta}.}
  \bibinfo{year}{2020}\natexlab{}.
\newblock \showarticletitle{Ember: energy management of batteryless event
  detection sensors with deep reinforcement learning}. In
  \bibinfo{booktitle}{\emph{Proceedings of the 18th Conference on Embedded
  Networked Sensor Systems}}. \bibinfo{pages}{503--516}.
\newblock


\bibitem[\protect\citeauthoryear{Fujita, Nagarajan, Kataoka, and
  Ishikawa}{Fujita et~al\mbox{.}}{2021}]%
        {pfrl}
\bibfield{author}{\bibinfo{person}{Yasuhiro Fujita}, \bibinfo{person}{Prabhat
  Nagarajan}, \bibinfo{person}{Toshiki Kataoka}, {and}
  \bibinfo{person}{Takahiro Ishikawa}.} \bibinfo{year}{2021}\natexlab{}.
\newblock \showarticletitle{ChainerRL: A Deep Reinforcement Learning Library}.
\newblock \bibinfo{journal}{\emph{Journal of Machine Learning Research}}
  \bibinfo{volume}{22}, \bibinfo{number}{77} (\bibinfo{year}{2021}),
  \bibinfo{pages}{1--14}.
\newblock
\urldef\tempurl%
\url{http://jmlr.org/papers/v22/20-376.html}
\showURL{%
\tempurl}


\bibitem[\protect\citeauthoryear{Hiremath, Yang, and Mankodiya}{Hiremath
  et~al\mbox{.}}{2014}]%
        {hiremath2014wearable}
\bibfield{author}{\bibinfo{person}{Shivayogi Hiremath}, \bibinfo{person}{Geng
  Yang}, {and} \bibinfo{person}{Kunal Mankodiya}.}
  \bibinfo{year}{2014}\natexlab{}.
\newblock \showarticletitle{Wearable Internet of Things: Concept, architectural
  components and promises for person-centered healthcare}. In
  \bibinfo{booktitle}{\emph{2014 4th International Conference on Wireless
  Mobile Communication and Healthcare-Transforming Healthcare Through
  Innovations in Mobile and Wireless Technologies (MOBIHEALTH)}}. IEEE,
  \bibinfo{pages}{304--307}.
\newblock


\bibitem[\protect\citeauthoryear{Kansal, Hsu, Zahedi, and Srivastava}{Kansal
  et~al\mbox{.}}{2007}]%
        {kansal2007power}
\bibfield{author}{\bibinfo{person}{Aman Kansal}, \bibinfo{person}{Jason Hsu},
  \bibinfo{person}{Sadaf Zahedi}, {and} \bibinfo{person}{Mani~B Srivastava}.}
  \bibinfo{year}{2007}\natexlab{}.
\newblock \showarticletitle{Power management in energy harvesting sensor
  networks}.
\newblock \bibinfo{journal}{\emph{ACM Transactions on Embedded Computing
  Systems (TECS)}} \bibinfo{volume}{6}, \bibinfo{number}{4}
  (\bibinfo{year}{2007}), \bibinfo{pages}{32--es}.
\newblock


\bibitem[\protect\citeauthoryear{Lau, Marakkalage, Zhou, Hassan, Yuen, Zhang,
  and Tan}{Lau et~al\mbox{.}}{2019}]%
        {lau2019survey}
\bibfield{author}{\bibinfo{person}{Billy Pik~Lik Lau},
  \bibinfo{person}{Sumudu~Hasala Marakkalage}, \bibinfo{person}{Yuren Zhou},
  \bibinfo{person}{Naveed~Ul Hassan}, \bibinfo{person}{Chau Yuen},
  \bibinfo{person}{Meng Zhang}, {and} \bibinfo{person}{U-Xuan Tan}.}
  \bibinfo{year}{2019}\natexlab{}.
\newblock \showarticletitle{A survey of data fusion in smart city
  applications}.
\newblock \bibinfo{journal}{\emph{Information Fusion}}  \bibinfo{volume}{52}
  (\bibinfo{year}{2019}), \bibinfo{pages}{357--374}.
\newblock


\bibitem[\protect\citeauthoryear{Schulman, Wolski, Dhariwal, Radford, and
  Klimov}{Schulman et~al\mbox{.}}{2017}]%
        {schulman2017proximal}
\bibfield{author}{\bibinfo{person}{John Schulman}, \bibinfo{person}{Filip
  Wolski}, \bibinfo{person}{Prafulla Dhariwal}, \bibinfo{person}{Alec Radford},
  {and} \bibinfo{person}{Oleg Klimov}.} \bibinfo{year}{2017}\natexlab{}.
\newblock \showarticletitle{Proximal policy optimization algorithms}.
\newblock \bibinfo{journal}{\emph{arXiv preprint arXiv:1707.06347}}
  (\bibinfo{year}{2017}).
\newblock


\bibitem[\protect\citeauthoryear{Tuncel, Bandyopadhyay, Kulshrestha, Mendez,
  and Ogras}{Tuncel et~al\mbox{.}}{2020}]%
        {tuncel2020towards}
\bibfield{author}{\bibinfo{person}{Yigit Tuncel}, \bibinfo{person}{Shiva
  Bandyopadhyay}, \bibinfo{person}{Shambhavi~V Kulshrestha},
  \bibinfo{person}{Audrey Mendez}, {and} \bibinfo{person}{Umit~Y Ogras}.}
  \bibinfo{year}{2020}\natexlab{}.
\newblock \showarticletitle{Towards wearable piezoelectric energy harvesting:
  Modeling and experimental validation}. In
  \bibinfo{booktitle}{\emph{Proceedings of the ACM/IEEE International Symposium
  on Low Power Electronics and Design}}. \bibinfo{pages}{55--60}.
\newblock


\bibitem[\protect\citeauthoryear{Tuncel, Basaklar, and Ogras}{Tuncel
  et~al\mbox{.}}{2021a}]%
        {tuncel2021much}
\bibfield{author}{\bibinfo{person}{Yigit Tuncel}, \bibinfo{person}{Toygun
  Basaklar}, {and} \bibinfo{person}{Umit Ogras}.}
  \bibinfo{year}{2021}\natexlab{a}.
\newblock \showarticletitle{How much energy can we harvest daily for wearable
  applications?}. In \bibinfo{booktitle}{\emph{2021 IEEE/ACM International
  Symposium on Low Power Electronics and Design (ISLPED)}}. IEEE,
  \bibinfo{pages}{1--6}.
\newblock


\bibitem[\protect\citeauthoryear{Tuncel, Bhat, Park, and Ogras}{Tuncel
  et~al\mbox{.}}{2021b}]%
        {tuncel2021eco}
\bibfield{author}{\bibinfo{person}{Yigit Tuncel}, \bibinfo{person}{Ganapati
  Bhat}, \bibinfo{person}{Jaehyun Park}, {and} \bibinfo{person}{Umit Ogras}.}
  \bibinfo{year}{2021}\natexlab{b}.
\newblock \showarticletitle{ECO: Enabling Energy-Neutral IoT Devices through
  Runtime Allocation of Harvested Energy}.
\newblock \bibinfo{journal}{\emph{IEEE Internet of Things Journal}}
  (\bibinfo{year}{2021}).
\newblock


\bibitem[\protect\citeauthoryear{{US Department of Labor}}{{US Department of
  Labor}}{2018}]%
        {amtus}
\bibfield{author}{\bibinfo{person}{{US Department of Labor}}.}
  \bibinfo{year}{2018}\natexlab{}.
\newblock \bibinfo{title}{{American Time Use Survey}}.
\newblock
\newblock
\newblock
\shownote{\url{https://www.bls.gov/tus/}, accessed 1 March 2021.}


\bibitem[\protect\citeauthoryear{Wang, Han, Leung, Niyato, Yan, and Chen}{Wang
  et~al\mbox{.}}{2020}]%
        {wang2020convergence}
\bibfield{author}{\bibinfo{person}{Xiaofei Wang}, \bibinfo{person}{Yiwen Han},
  \bibinfo{person}{Victor~CM Leung}, \bibinfo{person}{Dusit Niyato},
  \bibinfo{person}{Xueqiang Yan}, {and} \bibinfo{person}{Xu Chen}.}
  \bibinfo{year}{2020}\natexlab{}.
\newblock \showarticletitle{Convergence of edge computing and deep learning: A
  comprehensive survey}.
\newblock \bibinfo{journal}{\emph{IEEE Communications Surveys \& Tutorials}}
  \bibinfo{volume}{22}, \bibinfo{number}{2} (\bibinfo{year}{2020}),
  \bibinfo{pages}{869--904}.
\newblock


\bibitem[\protect\citeauthoryear{Yamin and Bhat}{Yamin and Bhat}{2021}]%
        {yamin2021online}
\bibfield{author}{\bibinfo{person}{Nuzhat Yamin} {and}
  \bibinfo{person}{Ganapati Bhat}.} \bibinfo{year}{2021}\natexlab{}.
\newblock \showarticletitle{Online solar energy prediction for
  energy-harvesting internet of things devices}. In
  \bibinfo{booktitle}{\emph{2021 IEEE/ACM International Symposium on Low Power
  Electronics and Design (ISLPED)}}. IEEE, \bibinfo{pages}{1--6}.
\newblock


\end{thebibliography}

% \appendix
% \renewcommand\thefigure{A.\arabic{figure}}
% \renewcommand{\theequation}{ A.\arabic{equation}}
% \input{files/appendix}

\end{document}